\documentclass[prd,aps,twocolumn,nofootinbib,showpacs,superscriptaddress]{revtex4-1}
\usepackage{amsfonts}
\usepackage{amsmath}
\usepackage{amssymb}
\usepackage{bm}
\usepackage{dcolumn}
\usepackage[dvips]{graphicx}
\usepackage{graphics}
\usepackage[latin1]{inputenc}
\usepackage{latexsym}
\usepackage{rotating}
\usepackage[colorlinks=true]{hyperref}
\usepackage{xspace} 
\usepackage[usenames]{color}
\usepackage{mathrsfs}
\usepackage{multirow}
\usepackage{pifont}
\usepackage{enumitem}

\definecolor {darkgreen}{rgb}{0.2,0.7,0.2}
\definecolor{purple}{rgb}{0.5,0,0.5}

\newcommand\be{\begin{equation}}
\newcommand\ba{\begin{eqnarray}}
\newcommand\ee{\end{equation}}
\newcommand\ea{\end{eqnarray}}

\newcommand\bw{\begin{widetext}}
\newcommand\ew{\end{widetext}}

\newcommand{\nn}{\nonumber}

\newcommand{\mrm}{\mathrm}

\begin{document}

\title{Improved Analytic Love-C Relations for Neutron Stars}

\author{Tristen Lowrey}
\affiliation{Department of Physics, University of Virginia, Charlottesville, Virginia 22904, USA}

\author{Kent Yagi}
\affiliation{Department of Physics, University of Virginia, Charlottesville, Virginia 22904, USA}

\author{Nicol\'as Yunes}
\affiliation{Illinois Center for Advanced Studies of the Universe, 
Department of Physics, 
University of Illinois Urbana-Champaign, 
Urbana, IL 61801, USA}

\begin{abstract} 
Precise measurements of neutron star observables (such as mass and radius) allow one to constrain the equations of state for supranuclear matter and develop a stronger understanding of nuclear physics. The Neutron star Interior Composition ExploreR (NICER) tracks X-ray hotspots on rotating NSs and is able to infer precise information about the compactness of the star. Gravitational waves
carry information about the tidal deformability (related to the tidal Love number) of neutron stars, which has been measured by the LIGO/Virgo/KAGRA collaboration. These two observables enjoy an approximately universal property between each other that is insensitive to the equations of state (the ``universal Love-C relation''). In this paper, we focus on deriving two analytic expressions for the Love-C relations that are ready-to-use and improve upon previous analytic expressions. The first model is inspired by a Newtonian polytrope, whose perturbation to the gravitational potential can be found analytically. We extend this Newtonian model to the relativistic regime by providing a quadratic fit to the gravitational potential perturbation against stellar compactness. 
The second model makes use of the Tolman VII model and adopts a spectral expansion with Chebyshev polynomials, which converges faster than the Taylor expansions used in previous work. We find that the first model provides a more accurate description of the Love-C relation for realistic neutron stars than the second model, and it provides the best expression among all other analytic relations studied here in terms of describing the averaged numerical Love-C relation. These new models are not only useful in practice, but they also show the power and importance of analytic modeling of neutron stars.
\end{abstract}

\maketitle

\section{Introduction}
\label{sec:Intro}
 
The central density of neutron stars (NSs) can possibly reach  several times nuclear saturation density ($\epsilon=2.8 \times 10^{14}$ g/cm$^3$) \cite{Lattimer:2006xb,Tan:2020ics,Tan:2021ahl}. Such a dense environment is highly challenging to realize on Earth, so NSs provide a unique opportunity to research extreme matter. The particular properties of these stars can be understood using a specific equation of state (EoS), which dictates the relationship between energy density and pressure inside a NS. Though NSs are known to have radii between 10\textbf{--}15 km and masses between $1M_\odot$ to $\sim 3M_\odot$, precise measurements of these characteristics could allow researchers to constrain EoSs and develop a stronger understanding of nuclear physics \cite{Lattimer:2004pg,Tan:2021nat,MUSES:2023hyz,Mroczek:2023zxo}.

Constraints have already been placed on the EoS through recent NS observations. The Neutron star Interior Composition ExploreR (NICER) tracks X-ray hotspots on rotating NSs and has the ability to infer precise measurements for the NS compactness ($C = G M/c^2 R$) and mass $M$ (and thus, the NS radius $R$)~\cite{10.1117/12.926396,10.1117/12.2056811}. 
Because the relation between mass and radius depends so sensitively on the underlying EoS, the results from NICER's observation of PSR J0030+0451~\cite{Riley:2019yda,Miller:2019cac} and PSR J0740+6620~\cite{Miller:2021qha,Riley:2021pdl} placed stringent bounds on the valid EoS~\cite{Raaijmakers:2019qny,Annala:2021gom,Pang:2021jta,Legred:2021hdx}. Gravitational waves (GWs) carry information about the interior structure of NSs through the (dimensionless) tidal deformability, $\Lambda$, which encodes  the susceptibility of the NS to be deformed by an external tidal field~\cite{Hinderer:2007mb,Chatziioannou:2020pqz}. A NS merger event detected by the LIGO/Virgo/KAGRA (LVK) collaboration, GW170817, measured this tidal deformability to $\Lambda= 190^{+390}_{-120}$ \cite{LIGOScientific:2018cki} for a $1.4M_\odot$ NS, which also places constraints on the EoS (see e.g.~\cite{Chatziioannou:2020pqz,LIGOScientific:2018cki,Annala:2017llu,Raithel:2018ncd,Lim:2018bkq,Bauswein:2017vtn,De:2018uhw,Most:2018hfd,Annala:2019puf,Malik:2018zcf,PhysRevD.99.043010,Carson:2019xxz,Raithel:2019ejc}). NICER and LVK constraints can in fact be combined to place joint bounds on the NS EoS~\cite{Raaijmakers:2019dks,Zimmerman:2020eho,Jiang:2019rcw,Dietrich:2020efo}.

Certain NS observables enjoy approximately universal relations that do not depend sensitively on the underlying EoS~\cite{Yagi:2016bkt,Doneva:2017jop}. 
One well-studied example is the relation among the moment of inertia ($I$), the tidal deformability or the Love number (Love), and the spin-induced quadrupole moment ($Q$), the so-called I-Love-Q relation~\cite{Yagi:2013awa,Yagi:2013bca}. 
In this paper, we will study, in particular, the approximately universal relation that exists between the tidal deformability and the compactness (the Love-C relation) of the star~\cite{Maselli:2013mva,Yagi:2016bkt}. Both of these quantities are measured independently through X-ray and GW observations, unlike the moment of inertia or quadrupole moment that has yet to be measured directly. Moreover, the Love-C relation has already been used to probe nuclear~\cite{Dong:2024opz} and gravitational~\cite{Silva:2020acr,Saffer:2021gak} physics.

Despite the complexity of this problem, there are analytic models that seek to capture the Love-C relation with varying levels of precision. One such model was presented by Zhao and Lattimer~\cite{Zhao:2018nyf}. They found that the tidal deformability is proportional to the compactness inversely raised to the sixth power, $C^{-6}$ for a certain range of stellar masses. Another model, offered by Jiang and Yagi~\cite{Jiang:2019vmf,Jiang:2020uvb}, uses the Tolman VII model to produce analytic expressions for various realistic NS profiles (energy density, pressure, and gravitational potential). The authors first improved the original Tolman VII model by introducing a parameter in the energy density expression to more accurately describe a realistic profile for the energy density. The authors then considered a tidal perturbation to the now improved Tolman VII model, and solved these equations by truncating a Taylor expansion in the stellar compactness and the dimensionless radial coordinate $\xi$. 
Furthermore, accurate fitting functions have also been constructed to reconstruct the Love-C behavior computed numerically with tabulated EoSs. A fit presented by two of the authors is one such example, and we include it here for future comparison~\cite{Yagi:2016bkt}. These analytic expressions for the Love-C relation offer not only ready-to-use expressions when analyzing data for NS observations, but also help understand the origin of the approximate universality~\cite{Jiang:2020uvb}.

In this paper, we seek to improve upon the aforementioned analytic models in two ways. In the first model, we derive an analytic relation for realistic EoSs inspired by Newtonian polytropes. That is, we first analytically solve a perturbation equation for the gravitational potential for an $n=1$ polytrope (where $n$ is the polytropic index) in the Newtonian limit. We then improve this analytic expression by creating a fit in one of the parameters as a function of stellar compactness (so that the analytic expression is valid even for high-compactness stars). In the second model, we adopt the Tolman VII model used by Jiang and Yagi~\cite{Jiang:2020uvb}, which is known to provide analytic expressions for realistic NS configurations. We here use a spectral expansion via Chebyshev functions~\cite{Chung:2023zdq,Chung:2023wkd,Chung:2024ira,Chung:2024vaf} that has a faster convergence than a Taylor expansion.

\begin{figure}
\includegraphics[width=8.5cm]{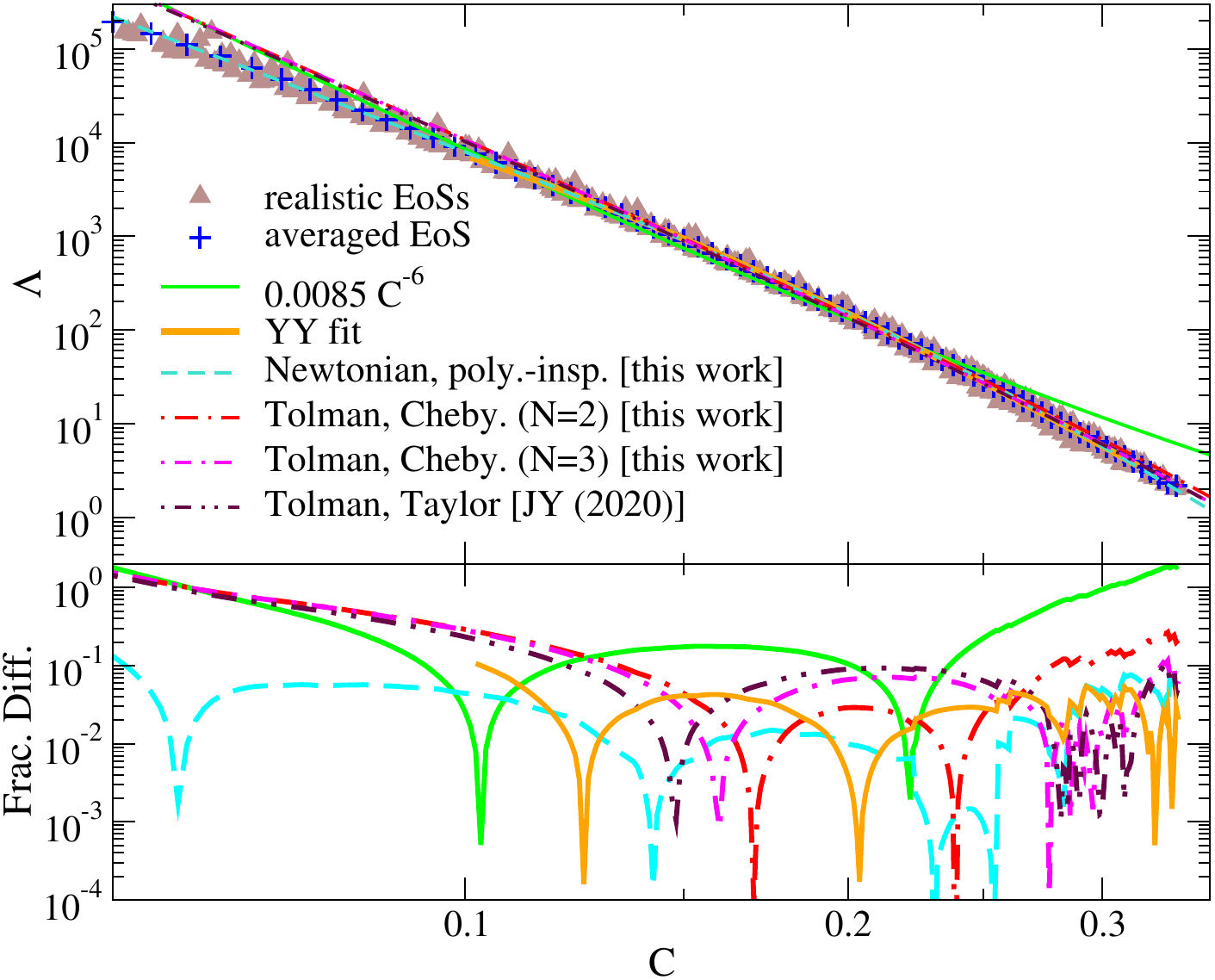} \\
\caption{\label{fig:summary}
(Top) Various Love-C relations for NSs. We present the two new analytic relations found in this paper: (i) the one inspired by a Newtonian $n=1$ polytrope that uses Eq.~\eqref{eq:y_poly_insp} for $y_*$ in Eq.~\eqref{eq:Lambda_C}
 (cyan dashed), and (ii) the one derived from the Tolman VII model with a Chebyshev expansion that uses Eq.~\eqref{eq:y-Tol-Cheby} for $y_*$ in Eq.~\eqref{eq:Lambda_C}, valid to second order (red dot-dashed) and third order (magenta dot-dashed) in compactness for the $y_*$ expression. We also present analytic relations for the Tolman VII model obtained via a Taylor expansion~\cite{Jiang:2020uvb} (brown dot-dashed), a relation $\Lambda \propto C^{-6}$ proposed in~\cite{Zhao:2018nyf} (green thin solid), and a quadratic fit in~\cite{Yagi:2016bkt} valid for $C > 0.1$ (orange thick solid). We also show the relations for realistic NSs found numerically using 27 EoSs for nuclear matter in~\cite{Yagi:2016bkt} (brown triangles), as well as the averaged relation among these 27 EoSs (blue crosses).
(Bottom) 
Fractional difference for each Love-C relation from the average EoS.
}
\end{figure}

\subsection*{Executive Summary}
\label{eq:summary}

For convenience, let us here summarize our main results. The dimensionless tidal deformability $\Lambda$ is given in terms of the compactness $C$ as~\cite{Hinderer:2007mb}
\begin{align}
\label{eq:Lambda_C}
\Lambda=& \frac{16}{15}(1-2 C)^2[2+2 C(y_*-1)-y_*] \nn \\
& \times\left\{2 C(6-3 y_*+3 C(5 y_*-8)) \right. \nn \\ 
& \left. +4 C^3\left[13-11 y_*+C(3 y_*-2)+2 C^2(1+y_*)\right]\right. \nn\\
&\left.+3(1-2 C)^2[2-y_*+2 C(y_*-1)] \log (1-2 C)\right\}^{-1}\,,
\end{align}
where $y_* = (rh'(r)/h(r))|_{r=R}$ and $h(r)$ is the radial part of the tidal perturbation to the gravitational potential.
We found two approximate expressions for $y_*$ in terms of $C$:
\begin{enumerate}
    \item \emph{Analytic expression inspired by Newtonian polytropes}:
\begin{align}
\label{eq:y_poly_insp}
y_*^\mathrm{N} = & -3 + \frac{2\sqrt{\bar \alpha \pi} J_1(2\sqrt{\bar \alpha \pi} )}{J_2(2\sqrt{\bar \alpha \pi} )} \nonumber \\
 = & -3 + \frac{4}{3} \bar \alpha \pi    \left(1-\frac{4 \bar \alpha  \pi 
   }{4 \bar \alpha \pi   +6 \sqrt{\bar \alpha \pi} 
   \cot \left(2 \sqrt{\bar \alpha \pi} \right)-3}\right)\,.
\end{align}
where $J_n$ is the spherical Bessel function of the first kind, while $\bar \alpha$ is given through fits as 
\begin{equation}
\label{eq:a_fit_summary}
    \bar \alpha^\mrm{fit} = c_0 +c_1 C + c_2 C^2\,,
\end{equation}
with
\begin{equation}
\label{eq:fit_coeff}
    (c_0,c_1,c_2) = (0.247,5.93,-17.7)\,.
\end{equation}

    \item \emph{Analytic expression via Tolman VII model}:
\begin{equation}
y_* = \sum_{j=0}^N \sum_{k=0}^K a_{j,k} \, T_{k}(1) \,C^j\,,
\end{equation}
where $T_k(\xi)$ are Chebyshev polynomials, and the coefficients $a_{j,k}$ are provided in Table~\ref{Table:a_coeff} up to $N=6$ and $K=6$. For $N=3$ and $K=6$, $y_*$ is given by
\begin{align}
    \label{eq:y-Tol-Cheby}
y_* =& \frac{21826059}{637247290} + \frac{92120622}{54053147} C \nn \\
    & + \frac{202561478}{60539327} C^2+ \frac{334154607}{47115665} C^3\nn \\
    &  + \mathcal{O}(C^4)\,.
\end{align}
\end{enumerate}  
When one Taylor expands either of the two models about $C \ll 1$, one finds that $y_* \to y_{*,0} = {\rm{const}}$, and then $\Lambda \sim [(2 - y_{*,0})/(9 + 3 y_{*,0})] C^{-5}$, which proves mathematically that $\Lambda$ does not scale as $C^{-6}$ in the Newtonian limit as proposed in~\cite{Zhao:2018nyf}. 

Figure~\ref{fig:summary} presents the above two analytic Love-C relations, as well as other analytic expressions for this approximately universal relation~\cite{Jiang:2020uvb,Zhao:2018nyf, Yagi:2016bkt}. We also present the relation for realistic NSs found numerically using 27 EoSs in~\cite{Yagi:2016bkt}, as well as an averaged relation. This figure also shows the fractional difference of each analytic relation with respect to the averaged EoS case. Observe that the first analytic model (Newtonian, polytrope-inspired) provides the most accurate description of the averaged Love-C relation in most of the compactness regimes and outperforms many of the other relations found analytically. This new relation is even better than the YY fit in~\cite{Yagi:2016bkt} even though both relations use the same number of fitting coefficients and were fitted against the same EoS data sets. For the Tolman VII relation, the analytic Love-C relation converges to the one obtained via numerical integration as we include higher orders in the stellar compactness. However, because the Tolman VII model adequately approximates stars with stiffer EoS in the low-compactness regime and softer EoS in the high-compactness regime, the model truncated at second or third order in compactness (presented in Fig.~\ref{fig:summary}) agrees better with the averaged EoS case than the one with higher order terms (e.g. sixth order in compactness). On the other hand, the polytrope-inspired model has no preference for EoS stiffness and can accurately describe all we present here. This is because the model involves fitting coefficients given in Eq.~\eqref{eq:a_fit_summary} to accurately reproduce the average behavior of the Love-C data. Finally, we see that the $C^{-6}$ prescription of~\cite{Zhao:2018nyf} is not accurate in the Newtonian limit ($C \ll 1$) or even for the more massive stars ($C > 0.25$), being accurate only in a small region of compactnesses around $C \sim 0.15$. Our results, especially the first analytic model, provide the most accurate analytical Love-C expression and show the importance of analytic studies on NS modeling.

\subsection*{Organization}

The remaining structure of the paper will be as follows.  
In Sec.~\ref{sec:tidal_deform} we discuss how to derive the tidal deformability in terms of the compactness. Then, in Sec.~\ref{sec:poly-insp}, we review how the Love-C relation is found in the Newtonian limit for an $n=1$ polytropic EoS. From there, we consider realistic EoSs and construct a new analytic Love-C relation. Next, we pivot and focus Sec~\ref{sec:tolman} on the Tolman VII model. We start by introducing the model and previous modifications before presenting our new method utilizing Chebyshev functions. Sec.~\ref{sec:conclusions} then summarizes our findings and discusses future work. We use the geometric units of $c= 1 = G$ throughout. 

\section{Tidal Deformability}
\label{sec:tidal_deform}

We first review how to calculate the tidal deformability from first principle, following mostly~\cite{Hinderer:2007mb,Damour:2009vw}. The metric ansatz of the background spherically symmetric spacetime is
\begin{equation}
    ds^2 = -e^{\nu(r)} dt^2 + e^{\lambda(r)} dr^2 + r^2 d\Omega^2\,.
\end{equation}
The tidal metric perturbation in $g_{tt}$ is given by
\begin{equation}
\delta g_{tt} = -e^\nu h(r) Y_{2m}(\theta,\phi)\,,
\end{equation}
where $Y_{2m}(\theta,\phi)$ are the spherical harmonics at $\ell = 2$. The radial part of the perturbation $h(r)$
satisfies~\cite{Hinderer:2007mb}:
\begin{align}
\label{eq:h_eq}
& h^{\prime \prime}+\left\{\frac{2}{r}+e^\lambda\left[\frac{2 m}{r^2}+4 \pi r(p-\epsilon)\right]\right\} h^{\prime} \nonumber \\
&
-\left(\frac{6 e^\lambda}{r^2}-4 \pi \frac{\alpha}{R^2}\right)h=0\,,
\end{align}
with a prime representing a radial derivative, $\epsilon$ and $p$ being the energy density and pressure respectively, $m(r)$ related to $\lambda(r)$ by $e^{-\lambda(r)} = 1-2m(r)/r$, and the dimensionless function $\alpha(r)$ is defined by
\begin{equation}
\label{eq:alpha}
\alpha(r) \equiv \left[e^\lambda\left(5 \epsilon+9 p+\frac{\epsilon+p}{c_s^2}\right)-\frac{\nu^{\prime 2}}{4\pi}\right] R^2\,,
\end{equation}
with the square of the sound speed given by $c_s^2 = dp/d\epsilon$.

Alternatively, we can write Eq.~\eqref{eq:h_eq} in first-order form, namely as
\begin{align}
\label{eq:riccati}
    r y' + y^2 + F y - Q = 0\,,
\end{align}
where $y = r h'/h$ is clearly dimensionless, and we have defined the dimensionless functions 
\begin{align}
    F &:= \frac{1 + 4 \pi r^2 (p - \epsilon)}{1 - 2 m/r}\,,
    \\
    Q &:= 6 \left(1 - \frac{2 m}{r}\right)^{-1} - 4 \pi \frac{r^2}{R^2} \alpha\,.
\end{align}
The first-order form of the tidal deformability equation makes it clear that it is a Riccati equation, which has the benefit that, if we could solve it analytically and in closed-form, the tidal deformability would then be simply determined by the solution evaluated at the surface through Eq.~\eqref{eq:Lambda_C}. The obvious disadvantage is that it is a Riccati equation, and therefore, it is manifestly non-linear and no closed-form, analytic, exact solutions are known. On the flip side, of course, no closed-form, analytic, exact solutions are known for the second-order linear form in Eq.~\eqref{eq:h_eq} either. As we shall see below, however, this Riccati form will yield some insight into the solution.

We now provide the expressions for the tidal Love number and (dimensionless) tidal deformability.
The tidal Love number is given by~\cite{Hinderer:2007mb}
\begin{align}
\label{eq:k2_eq}
k_2=& \frac{8 C^5}{5}(1-2 C)^2[2+2 C(y_*-1)-y_*] \nn \\
& \times\left\{2 C(6-3 y_*+3 C(5 y_*-8)) \right. \nn \\
& \left. +4 C^3\left[13-11 y_*+C(3 y_*-2)+2 C^2(1+y_*)\right]\right. \nn\\
&\left.+3(1-2 C)^2[2-y_*+2 C(y_*-1)] \log (1-2 C)\right\}^{-1}\,,
\end{align}
where recall that $C=M/R$ is the stellar compactness with mass $M$ and radius $R$, while $y_* := y(r=R)$. The dimensionless tidal deformability is then given by
\begin{equation}
\label{eq:Lambda_eq}
\Lambda = \frac{2}{3}\frac{k_2}{C^5}\,,
\end{equation} 
which is equivalent to Eq.~\eqref{eq:Lambda_C}. This quantity,
$\Lambda$, is related to the dimensionful tidal deformability $\lambda$ via $\Lambda = \lambda/M^5$, where $\lambda$ is given by the ratio between the tidally-induced quadrupole moment and the external quadrupolar tidal field strength~\cite{Hinderer:2007mb}. In the next two sections, we will explain, in turn, how to derive the two  analytic expressions for the approximately universal relation between $\Lambda$ and $C$ summarized in Sec.~\ref{eq:summary}.

\section{Analytic relation inspired by Newtonian polytropes}
\label{sec:poly-insp}

We first review the analytic relation inspired by Newtonian polytropes. We first show a fully-analytic relation for a specific polytropic EoS in the Newtonian limit. We next explain how to extend this to the relativistic case.

\subsection{Newtonian, $n=1$-Polytrope Love-C Relation}

To get some insight, let us first work in the Newtonian limit. Equation~\eqref{eq:h_eq} in this limit becomes
\begin{equation}
\label{eq:h_Newton}
h^{\prime \prime}+\frac{2}{r} h^{\prime}-\left(\frac{6}{r^2} - 4 \pi \frac{\alpha^\mathrm{N}(r)}{R^2}\right) h=0\,, \quad \alpha^\mathrm{N} (r)\equiv \frac{\epsilon}{c_s^2} R^2\,.
\end{equation}
As an example, we consider an $n=1$ polytropic EoS, $p = K \epsilon^2$, for which the speed of sound is $c_s^2 = 2 K \epsilon$. Using this and the relation between $K$ and $R$ for an $n=1$ polytrope (i.e.~$R = \sqrt{\pi K/2}$), we find $\alpha^\mathrm{N}$ to be a dimensionless constant: $\alpha^{\mathrm{N},(n=1)}= \pi/4$. 
In general, when $\alpha^\mathrm{N}$ is constant ($\bar \alpha$),  Eq.~\eqref{eq:h_Newton} becomes a Bessel equation and the solution that is regular at the center is given by~\cite{Hinderer:2007mb}
\begin{align}
h^\mathrm{N} \propto & J_2\left( \frac{2\sqrt{\bar \alpha \pi }r}{R} \right) \nonumber \\
\propto & \frac{4 \bar  \alpha  \pi  r^2-3 R^2}{r^3} \sin
   \left(\frac{2\sqrt{\bar \alpha \pi }r}{R} \right)+\frac{6\sqrt{\bar \alpha \pi } R}{r^2}
   \cos \left(\frac{2\sqrt{\bar \alpha \pi }r}{R} \right)\,,
\end{align}
where $J_n$ is the spherical Bessel function of the first kind. With this solution in hand, $y_*$ then becomes
\begin{align}
\label{eq:y_PN}
y_*^\mathrm{N} = & -3 + \frac{2\sqrt{\bar \alpha \pi} J_1(2\sqrt{\bar \alpha \pi} )}{J_2(2\sqrt{\bar \alpha \pi} )} \nonumber \\
 = & -3 + \frac{4}{3} \bar \alpha \pi    \left(1-\frac{4 \bar \alpha  \pi 
   }{4 \bar \alpha \pi   +6 \sqrt{\bar \alpha \pi} 
   \cot \left(2 \sqrt{\bar \alpha \pi} \right)-3}\right)\,.
\end{align}
For $n=1$ polytropes with $\bar \alpha^{(n=1)} = \pi/4$, the above $y_*^\mathrm{N}$ becomes
\begin{equation}
\label{eq:y-n1}
y_*^{\mathrm{N},(n=1)} = \frac{\pi^2-9}{3}\,.
\end{equation}
Inserting Eq.~\eqref{eq:y-n1} into Eqs.~\eqref{eq:k2_eq} and~\eqref{eq:Lambda_eq}, we find 
\allowdisplaybreaks[4]
\begin{widetext}
\begin{align}
\label{eq:Newtonian-n1}
\Lambda_{n=1} =& \frac{16}{15} (1-2 C)^2 \left[2 \left(\pi ^2-12\right) C-\pi ^2 +15\right] \left\{ 2C \left[\pi ^2 (C-1) (2 C-1)
[2 C (C+3)-3]  -3 C [2 C (C (4 C+11) \right. \right. \nn \\
& \left. -46)+69]+45\right]
\left. +3 \left[2 \left(\pi ^2-12\right)
C-\pi ^2+15\right] (1-2 C)^2 \log (1-2 C)\right\}^{-1}\,.
\end{align}
\end{widetext}
This analytic, approximate expression for the Love-C relation shall be referred to as the Newtonian, $n=1$-polytrope Love-C relation.

\begin{figure}
\includegraphics[width=8.5cm]{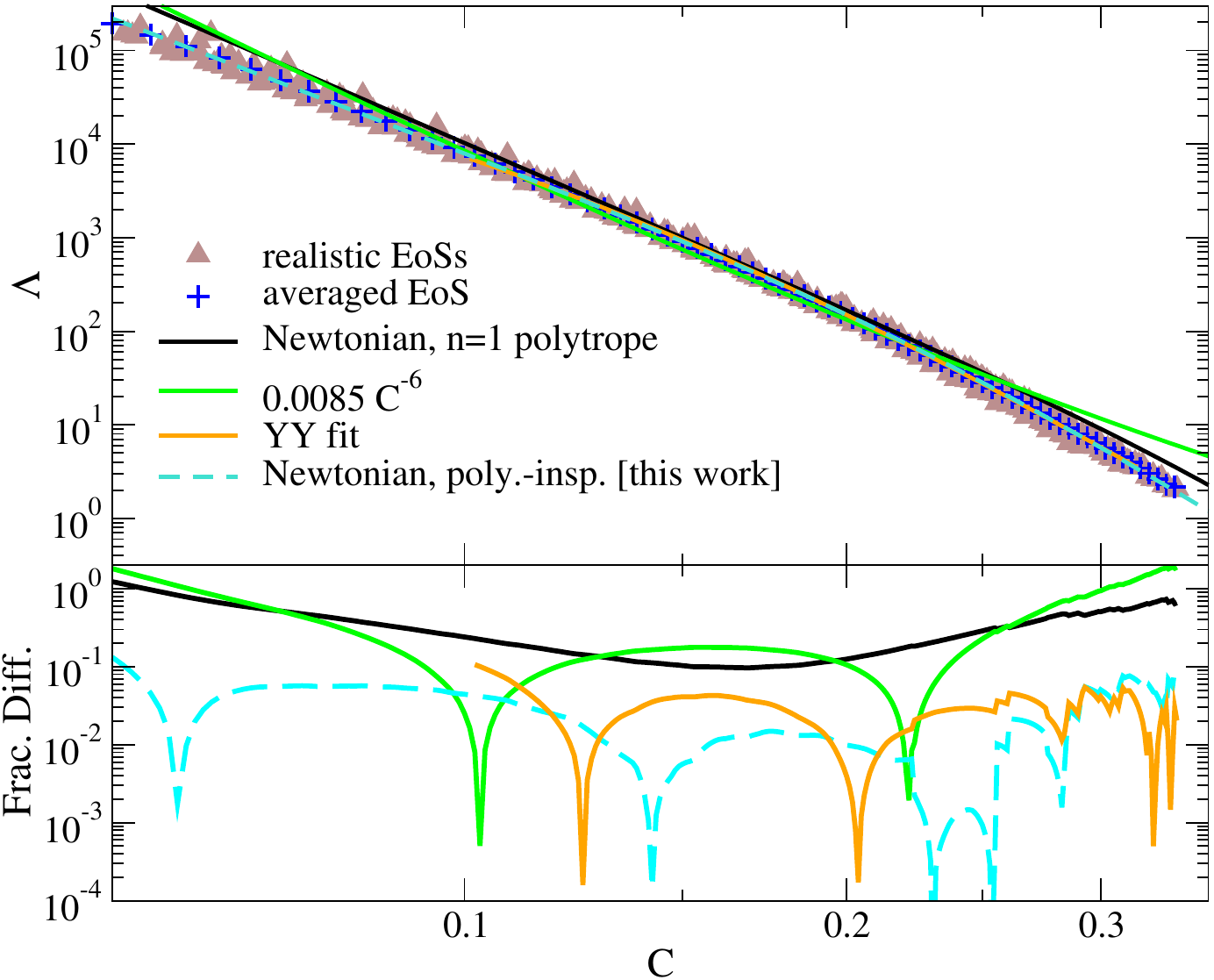}
\caption{\label{fig:LoveC-realistic}
Similar to Fig.~\ref{fig:summary} but for a reduced set of relations. We exclude the relations using the Chebyshev expansion but now include the one from Newtonian $n=1$ polytropes in Eq.~\eqref{eq:Newtonian-n1}. 
}
\end{figure}

The top panel of Fig.~\ref{fig:LoveC-realistic} compares this analytic relation with the Love-C relations for realistic EoSs. Here, we choose the same 27 nuclear EoSs considered in~\cite{Yagi:2016bkt} (excluding strange quark matter (SQM) EoSs). Some of these EoSs contain pion/kaon condensates and quark matter (see~\cite{Yagi:2016bkt,Chatziioannou:2015uea} for details on these EoSs). Among the Love-C relations for these EoSs, we compute an average $\Lambda$ for each $C$ to obtain the relation for this averaged EoS, i.e.
\begin{align}
\label{eq:ave_EoS}
    \Lambda_{\rm ave}(C) = \frac{1}{N} \sum_{i=1}^N \Lambda_i(C)\,,
\end{align}
where $\Lambda_i(C)$ is the Love-C relation for the $i$th EoS in the set of $N$ EoSs considered here\footnote{The number of EoSs used in our analysis is $N=27$ in most compactness regimes, though this number varies at high compactnesses, where the Love-C relation for some of the EoSs terminate after reaching their maximum compactness.}. The bottom panel of Fig.~\ref{fig:LoveC-realistic} presents the fractional difference between the analytic relations found above and the averaged Love-C relation. 
Although $\Lambda_{n=1}$ captures the qualitative behavior of the numerical relation for realistic EoSs, the quantitative agreement is rather poor. For comparison, we present the relation $\Lambda \propto C^{-6}$ proposed in~\cite{Zhao:2018nyf} as well as the relation from \cite{Yagi:2016bkt},
\begin{equation}
\label{eq:YYfit}
    C = \sum_{k=0}^{2} a_k^{YY} \Lambda^k\,,
\end{equation}
with $(a_0^{YY}, a_1^{YY}, a_2^{YY}) = (0.360, -0.0355, 0.000705)$.
Observe that the Love-C relation from YY is a more accurate description than the Newtonian, $n=1$ polytrope Love-C relation.
The simple relation $\Lambda \propto C^{-6}$ is a good model only for NSs with $C \in (0.1,0.2)$, as claimed in~\cite{Zhao:2018nyf}, but, in fact, it is not necessarily better than the Newtonian, $n=1$ polytrope Love-relation found above (as can be seen from the bottom panel of Fig.~\ref{fig:LoveC-realistic} when comparing the green and black curves).

\subsection{Newtonian, Polytrope-Inspired Love-C relation}

\begin{figure}
\includegraphics[width=8.5cm]{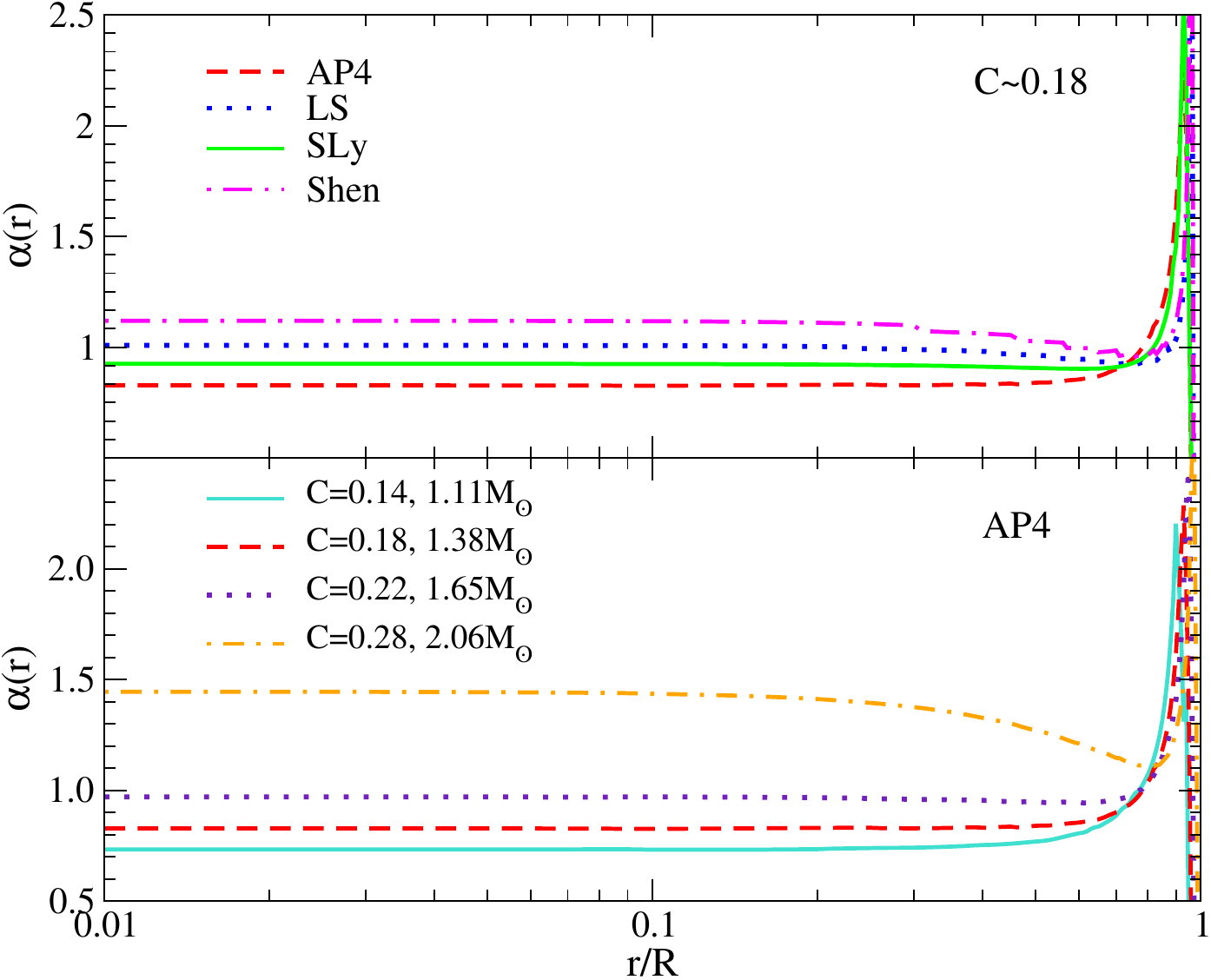}
\caption{\label{fig:alpha_profile}
(Top) $\alpha(r)$ against $r/R$ with $C\sim 0.18$ for selected EoSs. The corresponding mass of the NS is $1.38M_\odot$ (AP4), $1.39M_\odot$ (SLy), $1.51M_\odot$ (LS), and $1.86M_\odot$ (Shen). (Bottom) Similar to the top panel but for AP4 with different compactness/mass.
}
\end{figure}

Let us now study more realistic EoSs, and begin by studying the behavior of $\alpha(r)$. The top panel of Fig.~\ref{fig:alpha_profile} shows how $\alpha(r)$ evolves throughout the interior of a star for four example EoSs. We see that $\alpha(r)$ remains mostly constant for realistic NSs, particularly in the core region of the star, $r \lesssim 0.5 R$. This suggests we should use Eq.~\eqref{eq:y_PN} for $y_*$ even for realistic NSs, because this equation holds generically for Newtonian stars with $\alpha$ set to a constant.

\begin{figure}
\includegraphics[width=8.5cm]{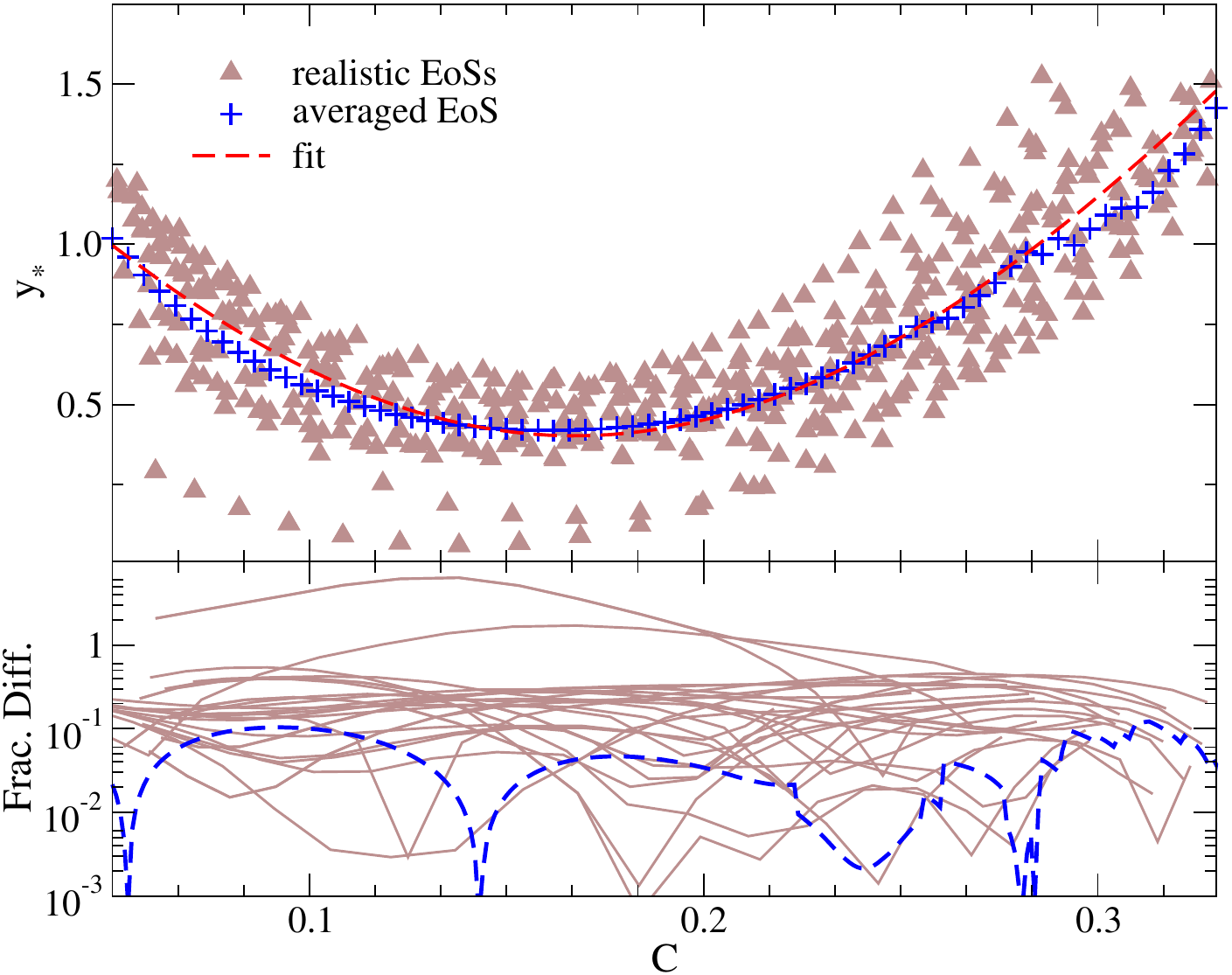}
\caption{\label{fig:yCfit}
(Top) Relation between $y_*$ and $C$ for the 27 realistic EoSs, averaged EoS, and the fit in Eqs.~\eqref{eq:y_poly_insp} and~\eqref{eq:a_fit_summary}. (Bottom) Fractional difference for each $y_*$--C relation from the fit. The fractional difference between the fit and the averaged EoS is shown with the blue dashed curve.
}
\end{figure}

The bottom panel of Fig.~\ref{fig:alpha_profile}, however, shows that this constant value of $\alpha$ not only varies with the EoS, but also with the compactness. Indeed, we see that although $\alpha$ remains almost constant in the core region for various $C$, the constant value varies depending on the EoS by about 20\%. Given this, one improvement we need to study is to account for this compactness dependence, so that the resulting tidal deformability is valid even for high-compactness stars. Empirically, we have found that the quadratic fitting function in Eq.~\eqref{eq:a_fit_summary} works well to accomplish this task\footnote{We also tested a number of other functional forms, including a cubic polynomial, a polynomial with the exponent being a fitting parameter, sinusoids, and hyperbolic functions. Among these fitting functions, the quadratic one was best at representing the averaged EoS.}. We determined the fitting coefficients of Eq.~\eqref{eq:fit_coeff} by fitting Eq.~\eqref{eq:a_fit_summary} against the combined data set for $y_*$ versus $C$ for 27 EoSs, as shown in Fig.~\ref{fig:yCfit}. Observe that the EoS variation in this relation is quite large, but the fit captures well the relation for the averaged EoS that was obtained similarly to the Love-C case in Eq.~\eqref{eq:ave_EoS}. We then inserted this fit for $y_*$ to Eq.~\eqref{eq:Lambda_C} to find the Love-C relation inspired by the Newtonian, $n=1$ polytropes.

The fit discussed above was not constructed by fitting against the numerical data for $\alpha$ versus $C$. One reason for this is that $\alpha$ is really a function of $r$, so the fit will depend on how we average $\alpha$ to find a constant value, which we then adopt as $\bar \alpha$. Another reason is that even if we had a perfect fit for $\bar \alpha$ against $C$, this fit would still not give us a good model for the Love-C relation. This is because the functional form for $y_*$ that we use in Eq.~\eqref{eq:y_PN} comes from the Newtonian limit, ignoring higher compactness corrections. We can avoid these issues by fitting against the $y_*$--C data directly. We compare the fit in Eq.~\eqref{eq:a_fit_summary} with the analytic relation for the Tolman VII model and numerical data in Appendix~\ref{app:alpha-C}.

Figure~\ref{fig:LoveC-realistic} also presents the Love-C relation for the updated analytic expression inspired by the Newtonian $n=1$ polytropes (``Newtonian, poly.-insp.''), as well as the fractional difference from the averaged EoS.
Observe that the new relation found here for realistic EoSs is reliable and accurate to $\mathcal{O}(1\%)$ when $0.1 \lesssim C \lesssim 0.3$. The new relation also has the smallest maximum fractional difference among all the analytic relations considered here. The new relation outperforms $\Lambda \propto C^{-6}$ and the Newtonian $n=1$ relation, and it describes the averaged EoS more accurately than the YY fit in most compactness regimes. Both the new relation and the YY fit use the same number of fitting coefficients and are fitted against the same 27 EoSs.

\section{Tolman VII Model}
\label{sec:tolman}

In this section, we revisit the Tolman VII model and find an approximate, analytic Love-C relation within this model. The model assumes the energy density profile for a spherically-symmetric NS takes the form~\cite{Tolman:1939jz}
\begin{equation}
    \epsilon \propto \left(1-\frac{r^2}{R^2}\right)\,.
\end{equation}
The background NS solution can be found in e.g.~\cite{Jiang:2019vmf}, and it has been seen to accurately describe realistic unperturbed NSs~\cite{Jiang:2019vmf}.
 
We next consider tidal perturbation to the Tolman VII model. Unlike the background Tolman VII solution, there is no known analytic solution to the tidal perturbation equation.  
Instead, we consider series expanding $y$ about small compactness:
\be
\label{eq:y_exp}
y(\xi, C) \simeq \sum_{j=0}^N y_j(\xi)C^j\,,
\ee
where $\xi= r/R$ is the dimensionless radial coordinate, with $R$ the stellar radius. 
After substituting this expansion and the background Tolman VII solution into the Riccati equation of Eq.~\eqref{eq:riccati}, we Taylor-expand the equation about $C=0$ and solve the resulting equations order by order in compactness. 

Let us discuss the leading $(j=0)$ case in more detail. The relevant differential equation for $y$ is found to be
\be
\label{eq:Tol0_LHS}
\xi\, y_0' + y_0 + y_0^2 - \frac{210-336\xi^2}{35-21\xi^2}=0 \,,
\ee
where a prime denotes a derivative with $\xi$. This equation can be solved analytically and the solution is given in terms of a hypergeometric function. The solution $y_0$ evaluated at the stellar radius $(\xi=1)$ is then given by
\begin{equation}
\label{eq:y_Tol}
    y_*^\mrm{(Tol)} = 2-\frac{6 \, _2F_1\left(\frac{1}{4} \left(9-\sqrt{65}\right),\frac{1}{4}
   \left(9+\sqrt{65}\right);\frac{9}{2};\frac{3}{5}\right)}{7 \,
   _2F_1\left(\frac{1}{4} \left(5-\sqrt{65}\right),\frac{1}{4}
   \left(5+\sqrt{65}\right);\frac{7}{2};\frac{3}{5}\right)}\,.
\end{equation}

Unfortunately, the equations can be solved analytically for $y_j$ only at this leading order in $C$. Jiang and Yagi~\cite{Jiang:2020uvb} thus took an approximate approach by Taylor expanding $y_{j>0} (\xi)$ about $\xi = 0$:
\be
y_j(\xi) \simeq \sum_{k=1}^K c_{j}^{(2k)} \, \xi^{2k}\,.
\ee
Jiang and Yagi studied this representation to 12th order in $\xi$ ($K=6$) and 6th order in $C$ ($N=6$) to find that the most accurate results were obtained when keeping terms up to $\mathcal{O}(C^3)$. In this case, $y_*$ is given by~\cite{Jiang:2020uvb} 
\begin{align}
    y_*^\mrm{(Tol,JY)} =& \frac{75974923394}{756262478125}+\frac{6131587821173
   }{3932564886250}C \nonumber \\
&+\frac{68922705930941}{20819461162500}C^2 \nn \\
&+\frac{107518543893319087}{14865095270025000}C^3+\mathcal{O}(C^4)\,.
\end{align}

Because the convergence for a Taylor expansion is rather slow, let us here,
instead, expand $y_j(\xi)$ in terms of Chebyshev polynomials $T_k(\xi)$:
\be
\label{eq:yn_exp}
y_j(\xi) \simeq \sum_{k=0}^K a_{j,k} \, T_{k}(\xi)\,,
\ee
where only even $k$ values are non-vanishing. A similar spectral expansion has been used in e.g.~\cite{Chung:2023zdq,Chung:2023wkd,Chung:2024ira,Chung:2024vaf}.
As before, let us first focus on the leading-order-in-compactness $(j=0)$ case, where the exact solution is given in Eq.~\eqref{eq:y_Tol}. Inserting Eq.~\eqref{eq:yn_exp} with $j=0$ and $K=6$ into Eq.~\eqref{eq:Tol0_LHS} yields
\begin{widetext}
\begin{align}
\label{eq:chebyriccati0}
&\xi \left[4a_{0,2} - a_{0,4}\left(16\xi-32\xi^3\right) +a_{0,6}\left(36\xi-192\xi^3+192\xi^5\right)\right]  \nonumber \\ 
&+ \left[a_{0,0}-a_{0,2}\left(1-2\xi^2\right)+a_{0,4}\left(1-8\xi^2+8\xi^4\right)
-a_{0,6}\left(1-18\xi^2+48\xi^4-32\xi^6\right)\right]^2 
\nonumber \\
&+ a_{0,0} - a_{0,2}\left(1-2\xi^2\right) +a_{0,4}\left(1-8\xi^2+8\xi^4\right) -a_{0,6}\left(1-18\xi^2+48\xi^4-32\xi^6\right) - \frac{210-336\xi^2}{35-21\xi^2} =0\,.
\end{align}
\end{widetext}
Using the Chebyshev normalization condition given by
\be
\label{eq:cheby_norm}
\int_{-1}^{1} T_n(C)T_m(C) \frac{\,dC}{\sqrt{1-C^2}} =
    \begin{cases}
        0 & \text{if } n\neq m\\
        \pi & \text{if } n=m=0\\
        \frac{\pi}{2} & \text{if } n=m\neq 0
    \end{cases}
\,,
\ee
one can construct a system of equations to solve for the $a_{0,n}$ coefficients. At zeroth-order, the presence of a $y_0^2(\xi)$ term allows for multiple sets of solutions, where one can determine the correct set by comparing it with the boundary condition $y_0(0)=2$. We present values for the $a_{0,k}$ coefficients in the first row of Table~\ref{Table:a_coeff}.
These coefficients can be obtained analytically, though the expressions are lengthy, so we only present here their numerical values.

\begin{table}
\begin{centering}
\begin{tabular}{cccccccccccccccccccccccccccc}
\hline
\hline
\noalign{\smallskip}
\multicolumn{1}{c}{$a_{0,0}$}
& \multicolumn{1}{c}{$a_{0,2}$} 
& \multicolumn{1}{c}{$a_{0,4}$} 
& \multicolumn{1}{c}{$a_{0,6}$} 
\\
\hline
\noalign{\smallskip}
\multicolumn{1}{c}{$1.21971$} & 
\multicolumn{1}{c}{$-0.938577$} & 
\multicolumn{1}{c}{$-0.189775$} &
\multicolumn{1}{c}{$-0.0571046$}
\\
\noalign{\smallskip}
\multicolumn{1}{c}{$a_{1,0}$}
& \multicolumn{1}{c}{$a_{1,2}$} 
& \multicolumn{1}{c}{$a_{1,4}$} 
& \multicolumn{1}{c}{$a_{1,6}$} 
\\
\hline
\noalign{\smallskip}
\multicolumn{1}{c}{$0.648441$} & 
\multicolumn{1}{c}{$0.773603$} & 
\multicolumn{1}{c}{$0.177212$} &
\multicolumn{1}{c}{$0.105004$} 
\\
\noalign{\smallskip}
\multicolumn{1}{c}{$a_{2,0}$}
& \multicolumn{1}{c}{$a_{2,2}$} 
& \multicolumn{1}{c}{$a_{2,4}$} 
& \multicolumn{1}{c}{$a_{2,6}$} 
\\
\hline
\noalign{\smallskip}
\multicolumn{1}{c}{$0.892301$} & 
\multicolumn{1}{c}{$1.75068$} & 
\multicolumn{1}{c}{$0.737539$} &
\multicolumn{1}{c}{$-0.034573$} 
\\
\noalign{\smallskip}
\multicolumn{1}{c}{$a_{3,0}$}
& \multicolumn{1}{c}{$a_{3,2}$} 
& \multicolumn{1}{c}{$a_{3,4}$} 
& \multicolumn{1}{c}{$a_{3,6}$} 
\\
\hline
\noalign{\smallskip}
\multicolumn{1}{c}{$1.54571$} & 
\multicolumn{1}{c}{$3.94102$} & 
\multicolumn{1}{c}{$2.09982$} &
\multicolumn{1}{c}{$-0.494328$} 
\\
\noalign{\smallskip}
\multicolumn{1}{c}{$a_{4,0}$}
& \multicolumn{1}{c}{$a_{4,2}$} 
& \multicolumn{1}{c}{$a_{4,4}$} 
& \multicolumn{1}{c}{$a_{4,6}$} 
\\
\hline
\noalign{\smallskip}
\multicolumn{1}{c}{$3.08069$} & 
\multicolumn{1}{c}{$9.04228$} & 
\multicolumn{1}{c}{$5.2485$} &
\multicolumn{1}{c}{$-1.85743$} 
\\
\noalign{\smallskip}
\multicolumn{1}{c}{$a_{5,0}$}
& \multicolumn{1}{c}{$a_{5,2}$} 
& \multicolumn{1}{c}{$a_{5,4}$} 
& \multicolumn{1}{c}{$a_{5,6}$} 
\\
\hline
\noalign{\smallskip}
\multicolumn{1}{c}{$6.66921$} & 
\multicolumn{1}{c}{$20.9726$} & 
\multicolumn{1}{c}{$12.3184$} &
\multicolumn{1}{c}{$-5.63809$} 
\\
\noalign{\smallskip}
\multicolumn{1}{c}{$a_{6,0}$}
& \multicolumn{1}{c}{$a_{6,2}$} 
& \multicolumn{1}{c}{$a_{6,4}$} 
& \multicolumn{1}{c}{$a_{6,6}$} 
\\
\hline
\noalign{\smallskip}
\multicolumn{1}{c}{$15.1007$} & 
\multicolumn{1}{c}{$48.8426$} & 
\multicolumn{1}{c}{$27.7903$} &
\multicolumn{1}{c}{$-15.6508$} 
\\
\noalign{\smallskip}
\hline
\hline
\end{tabular}
\end{centering}
\caption{\label{Table:a_coeff} Coefficients $a_{j,k}$ in Eq.~\eqref{eq:yn_exp} for $y_j$ as a function of the dimensionless radial coordinate  $\xi$ within the Chebyshev expansion.}
\end{table}

\begin{figure}
\includegraphics[width=8.5cm]{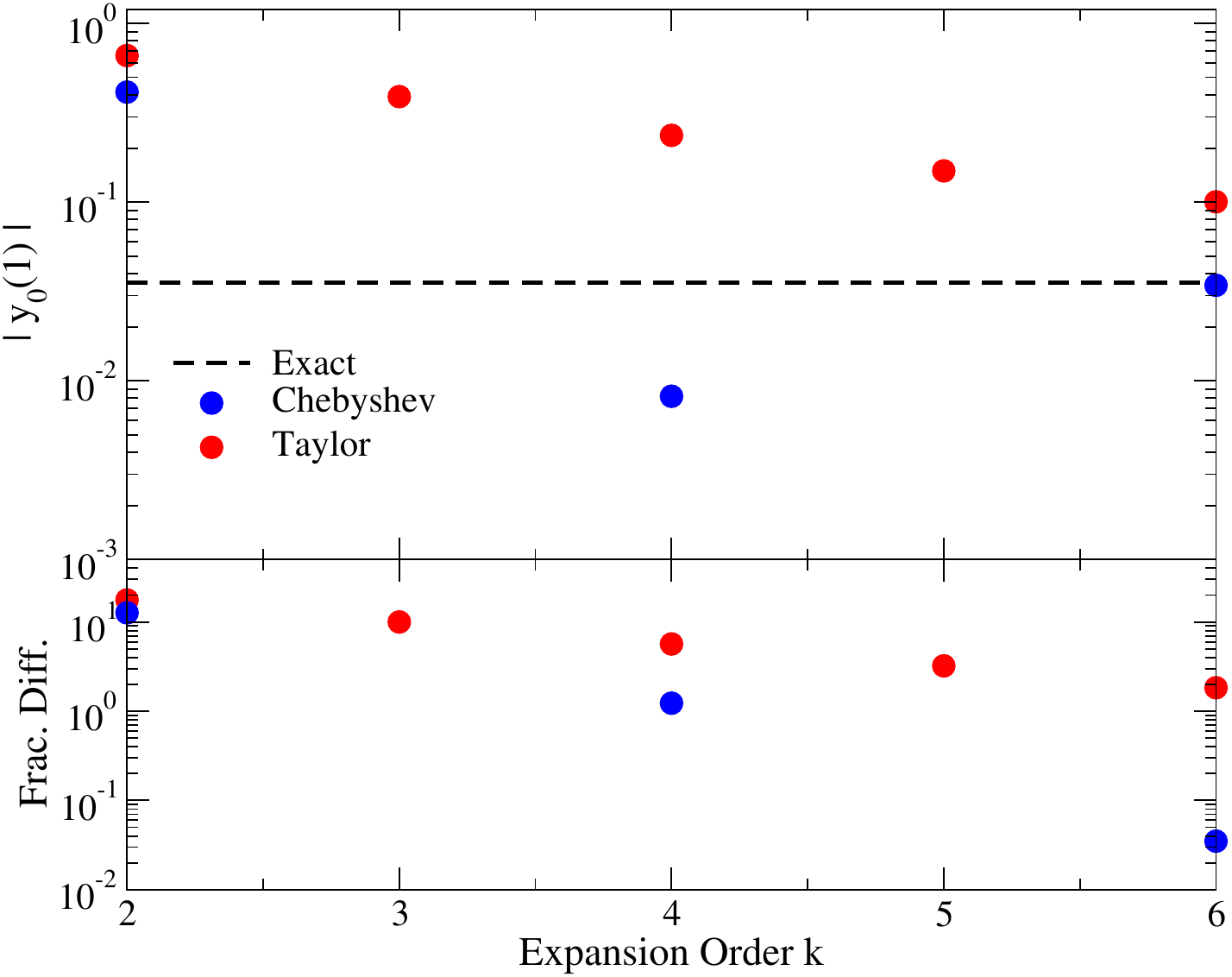}
\caption{\label{fig:expconv0}
(Top) $|y_0(1)|$ as a function of the expansion order $K$. We compare the Taylor expansion in~\cite{Jiang:2020uvb} (red dots) with the new Chebyshev expansion (blue dots) for orders 2, 4, and 6 against the exact Tolman value (black solid) computed from the hypergeometric series in Eq.~(\ref{eq:y_Tol}). (Bottom) Fractional difference of $|y_0(1)|$ from approximate expansions against the exact Tolman value as a function of the expansion orders. 
}
\end{figure}
\begin{figure}
\includegraphics[width=8.5cm]{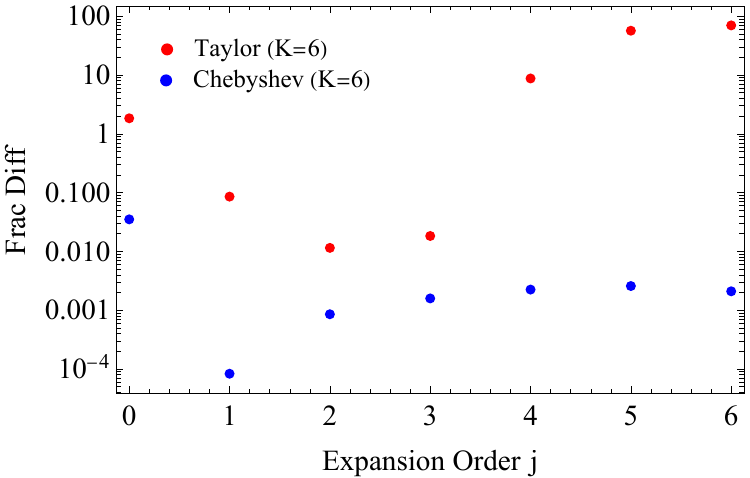}
\caption{\label{fig:expfracdiff_all}
Fractional difference of $y_j(1)$ at various compactness expansion orders $j$ for the Chebyshev and Taylor results with orders $K=6$ against the exact $(j=0)$ or numerical $(j>0)$ Tolman value.}
\end{figure}

\begin{figure}
\includegraphics[width=8.5cm]{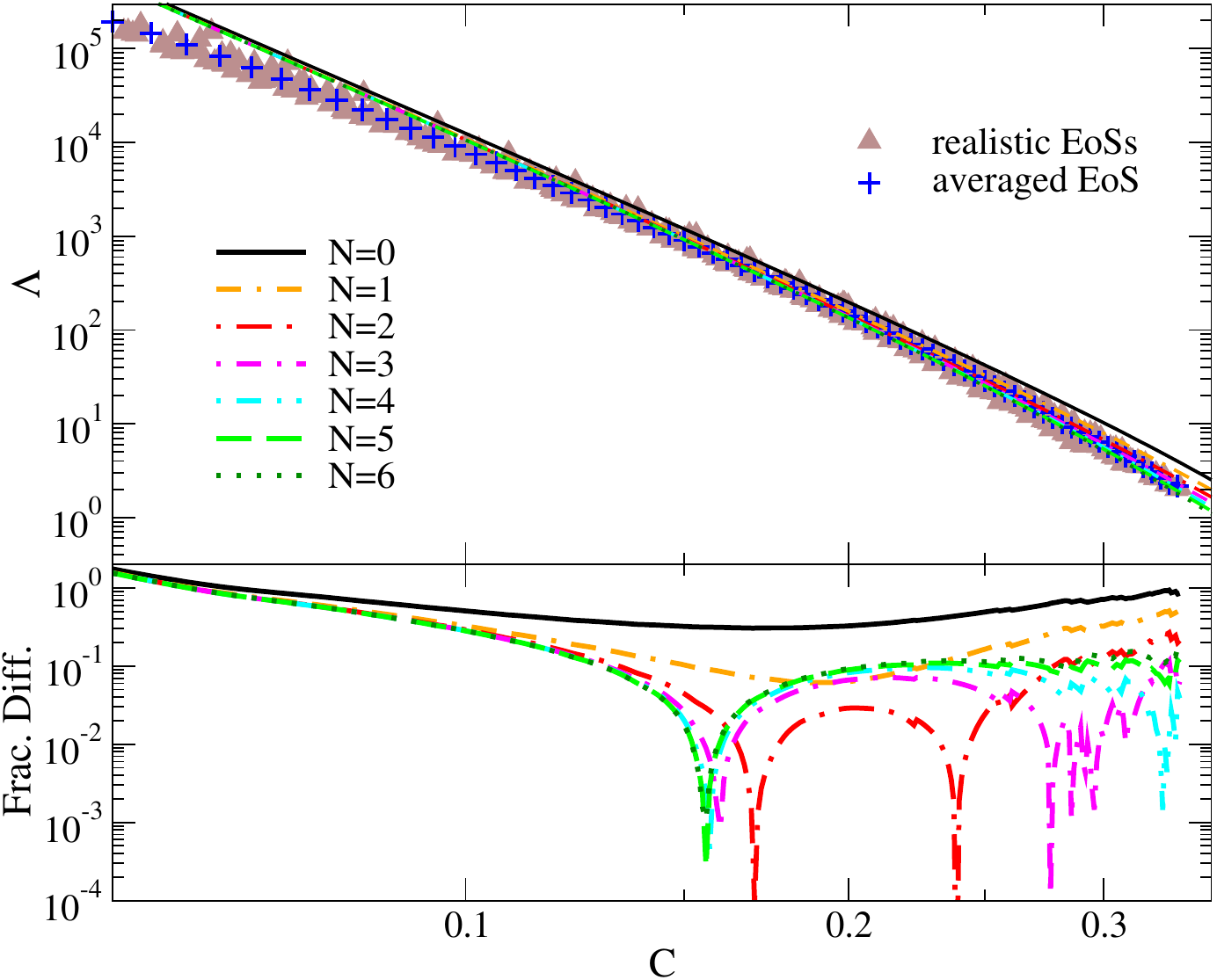}
\caption{\label{fig:chebyvsdata}
(Top) Analytic Chebyshev Love-C relations for orders $N \in [0,6]$ presented with the realistic EoSs and their average. (Bottom) Fractional difference between analytic Chebyshev Love-C relations for orders $N \in [0,6]$ against the average data set.  
}
\end{figure}

Let us now investigate how well the Chebyshev expansion can reproduce the true value for $y$ at the surface. 
The top panel of Fig.~\ref{fig:expconv0} compares $y_0(1)$ obtained  through the Chebyshev and Taylor approximations against the exact solution in Eq.~\eqref{eq:y_Tol} at different truncation order $K$ in the expansion. The bottom panel shows the fractional difference between each of the approximate values (with Chebyshev and Taylor expansions) and the exact one. 
Notice that the Chebyshev expansion converges very accurately by the 6th order while the Taylor expansion is only about as accurate at the 6th order as the Chebyshev one is at the 4th order. 

We can repeat the $j=0$ analysis to higher order in $j$ to find the Chebyshev coefficients $a_{j,k}$. Since we could not solve the Riccati equation analytically for $j\geq 1$ to check the accuracy of the Chebyshev expansion, we solved the equation numerically to find the numerical value of $y_j(1)$. We then compare this against those from the Chebyshev and Taylor expansions. 
Figure~\ref{fig:expfracdiff_all} presents the fractional difference in $y_j(1)$ for these  expansions relative to the exact ($j=0$) and the numerical ($j\geq  1$) values, as a function of different expansion orders in compactness ($j$). We use the expansion order of $K=6$ for the Chebyshev and Taylor expansions. Notice that the Taylor expansion becomes inaccurate after $j=4$, while the Chebyshev one remains accurate through $j=6$. 

Let us finally study how accurately the Love-C relation from the Chebyshev expansion can model the relation from realistic EoSs. The Love-C relation can be obtained by substituting $y_j(1)$ from the Chebyshev expansion into $y_*$ in Eq.~\eqref{eq:k2_eq}, which is further substituted into Eq.~\eqref{eq:Lambda_eq}. 
The top panel of Fig.~\ref{fig:chebyvsdata} shows how the Love-C relations at different compactness-expansion-orders $N$ converge from the Chebyshev expansion. We also present the relation from realistic equations of state and their average  for comparison. We see that the relations converge as we go to higher order $N$.

The bottom panel of Fig.~\ref{fig:chebyvsdata} shows the fractional difference between these Love-C relations at different Chebyshev expansion orders and the averaged EoS one. Observe that the Love-C relation from the Chebyshev expansion appears to have very little variability regardless of the expansion order at low compactness and can model the realistic EoSs well only in the high compactness regime. This is because the Tolman VII EoS corresponds to a stiff (EoS) model for low compactness, and it softens up at larger compactnesses. Notice also that as one increases in $N$, the relation converges quickly to a particular form, best seen at $N=6$, as we expect. 
Although the accuracy to correctly recover the Tolman VII model increases for higher order $N$, the $N=2$ or $N=3$ model describes most accurately the Love-C relation for the averaged EoS. The $y_*$ value in this case is given in Eq.~\eqref{eq:y-Tol-Cheby}.

Returning to Fig.~\ref{fig:summary}, we showed there the Chebyshev expansion result at $N=2$ and $N=3$ with other methods discussed earlier. We see that the Chebyshev relations work decently well in most of the intermediate and high compactness regimes, outperforming $\Lambda \propto C^{-6}$ and the Taylor-expanded Tolman expression in modeling the Love-C relation from realistic EoSs. However, the Newtonian polytrope inspired model is still the most accurate one overall, followed by the fit presented by two of the authors in~\cite{Yagi:2016bkt}.

\section{Conclusions}
\label{sec:conclusions}

In this paper, we derived and offered two new, analytic models for the approximately-universal Love-C relation of NSs, and found them both to be successful in different respects. 
The first model was inspired by an $n=1$ Newtonian polytrope. We extended this to a realistic NS by assuming $\alpha(r)$ is a constant and constructed a quadratic fit in terms of compactness. The resulting Love-C relation is reliable and accurate to $\mathcal{O}(1\%)$ for a wide range of compactnesses. This new analytic relation outperforms most other analytic relations considered here, and it is even more accurate than fits found previously, e.g.~in~\cite{Yagi:2016bkt}.

We also considered a modified Tolman VII model where, in the absence of an analytic solution to the tidal perturbation equation, we series expanded $y$ in terms of Chebyshev functions first about small compactness, and then about the dimensionless radius $\xi=0$. In comparison to the previous Taylor-expanded model, the Chebyshev expansion proved superior at recovering the exact/numerical Tolman value through sixth order in compactness. When comparing the analytic Chebyshev Love-C relations against other models, the former showed greater accuracy over the simple $\Lambda \propto C^{-6}$ relation  in the medium to high compactness regime. However, this expansion was not more accurate than the analytic relation first offered. This is, again, because the Newtonian polytrope-inspired one involves fitting on the averaged Love-C data using the coefficients given in Eq.~\eqref{eq:a_fit_summary}. As a result, the model has no such preference for stiff or soft EoSs.

We end this note by providing a few avenues for future work. In~\cite{Jiang:2019vmf,Jiang:2020uvb}, the authors constructed an improved Tolman VII model by introducing an extra parameter to the energy density profile to better model a realistic EoS profile. The authors were then able to show analytically the amount of EoS-variation in the Love-C relation, which is consistent with the $\mathcal{O}(10\%)$ variability that is usually found numerically. It would be interesting to extend our second analytic Love-C model (Tolman VII with Chebyshev expansion) to the improved Tolman VII one. This way, one should be able to find an analytic relation that better approximates the relation obtained using an averaged EoS. Another interesting avenue for future work is to consider EoSs with non-trivial features (e.g. bumps, oscillations, and plateaus) in the speed of sound, as motivated by the possible appearance of exotic degrees of freedom in extreme matter~\cite{Tan:2021nat,Mroczek:2023eff,Mroczek:2023zxo}, and to study how such features affect the Love-C approximate universality.

\acknowledgments

K.Y. acknowledges support from NSF Grant PHY-2207349, PHY-2309066, PHYS-2339969, and the Owens Family Foundation. N. Y. acknowledges support from the Simons Foundation through Award No. 896696, the National Science Foundation (NSF) Grant through No. PHY-2207650, and NASA through Grant No. 80NSSC22K0806.

\appendix

\section{Analytic relation between $\alpha$ and $C$}
\label{app:alpha-C}

In this appendix, we estimate analytically the relation between $\alpha$ and $C$ and compare it with the fit in Eq.~\eqref{eq:a_fit_summary}. We begin by working in the Newtonian limit. We first rewrite $\epsilon/c_s^2$ (which enters the definition of $\alpha^\mathrm{N}$ in Eq.~\eqref{eq:h_Newton}) as 
\begin{align}
    \frac{\epsilon}{c_s^2} = \epsilon \frac{d \epsilon}{dp} = \frac{1}{2} \frac{d  \epsilon^2}{dp}\,. 
\end{align}
Then, $\alpha^\mathrm{N}$ is simply
\begin{align}
    \alpha^\mathrm{N} = \frac{1}{2}\frac{M^2}{C^2}\frac{d  \epsilon^2}{dp}\,.
\end{align}

We next\footnote{We used the assumption that $\epsilon(R)=0$ when deriving Eq.~\eqref{eq:alpha_ave_p}, so this expression does not apply to e.g. incompressible stars with constant density (in that case, $c_s^2 = \infty$ and $\alpha^\mathrm{N}=0$). 
} consider taking the average of $\alpha^\mathrm{N}$ over $p$, with which one finds
\begin{align}
\label{eq:alpha_ave_p}
    \langle  \alpha^\mathrm{N} \rangle_p = \frac{\int_{p_c}^0 \alpha^\mathrm{N} dp}{\int_{p_c}^0 dp} = \frac{1}{2}\frac{M^2}{C^2} \frac{\epsilon_c^2}{p_c}\,,
\end{align}
where $p_c$ and $\epsilon_c$ are the central pressure and energy density. Saes \emph{et al}.~\cite{Saes:2024xmv} showed that $\langle c_s^2 \rangle_\epsilon = p_c/\epsilon_c$ when $c_s^2$ is integrated over the energy density, so we can further rewrite the above equation as
\begin{align}
\label{eq:alpha_ave_p_2}
    \langle  \alpha^\mathrm{N} \rangle_p = \frac{1}{2}\frac{M^2}{C^2} \frac{\epsilon_c}{\langle c_s^2 \rangle_\epsilon}\,.
\end{align}
We then see that the average value of $\alpha$ in the Newtonian limit seems to be related to the reciprocal of the speed of sound, averaged over densities, which was found in~\cite{Saes:2024xmv} to present universal behavior with the compactness.  

For a concrete example, let us now consider the Tolman VII model used in Sec.~\ref{sec:tolman}. To leading order in $C$, we have~\cite{Saes:2024xmv} 
\begin{align}
\label{eq:epsilon_cs_Tol}
    \epsilon_c = \frac{15 C}{8 \pi  R^2} + \mathcal{O}(C^2)\,, \quad   \langle c_s^2 \rangle_\epsilon = \frac{C}{2} + \mathcal{O}(C^2)\,.
\end{align}
Using these into Eq.~\eqref{eq:alpha_ave_p_2}, we find\footnote{ We can repeat the analysis for other analytic models, such as an $n=1$ polytrope of the form $p = K \epsilon^2$. In this case, $\epsilon_c = \pi C/(4 R^2) + \mathcal{O}(C^2)$ while $\langle c_s^2 \rangle_\epsilon$ is the same as the Tolman VII case in Eq.~\eqref{eq:epsilon_cs_Tol}, leading to
\begin{equation}
     \langle  \alpha^\mathrm{N, n=1} \rangle_p = \frac{\pi}{4}  + \mathcal{O}(C^2) = 0.785 + \mathcal{O}(C^2)\,.
\end{equation} 
}
\begin{align}
     \langle  \alpha^\mathrm{N, Tol} \rangle_p = \frac{15}{8 \pi}  + \mathcal{O}(C^2)\,.
\end{align}
We can find higher order corrections by using the Tolman VII solution in e.g.~\cite{Jiang:2019vmf} to Eq.~\eqref{eq:alpha} for $\alpha(r)$ and performing the integral
\begin{align}
    \langle  \alpha \rangle_p  =\frac{\int_{p_c}^0 \alpha dp}{\int_{p_c}^0 dp} = - \frac{1}{p_c}\int_{0}^R \alpha^\mathrm{N} \frac{dp}{dr}dr \,.
\end{align}
We then find
\begin{align}
\label{eq:alpha-C-Tol}
    \langle  \alpha \rangle_p = & \frac{15}{8 \pi }+\frac{167C}{32 \pi }+\frac{9437 C^2}{1344 \pi }+O\left(C^{3}\right) \nonumber \\
=&  0.597+1.66  C+2.24 
  C^2+O\left(C^{3}\right)\,.
\end{align}

\begin{figure}
\includegraphics[width=8.5cm]{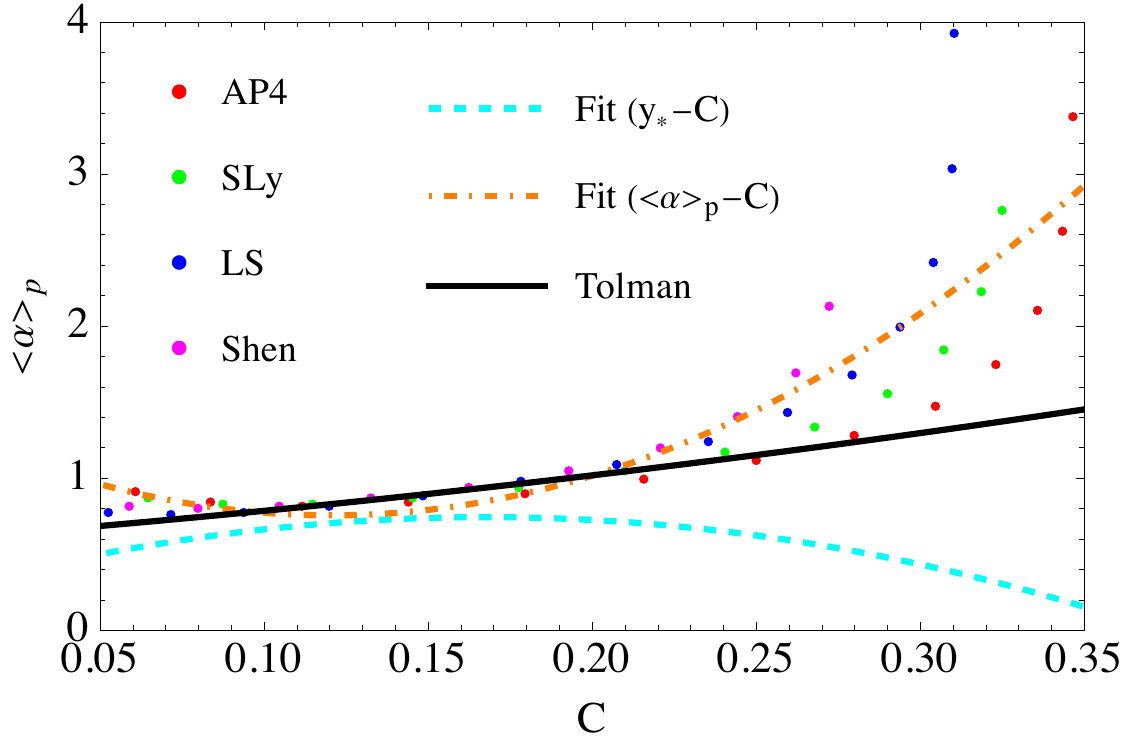}
\caption{\label{fig:alpha-C}
Relation between $\langle  \alpha \rangle_p$ and C for selected EoSs. We also present the analytic relation from the quadratic fits in Eq.~\eqref{eq:a_fit_summary} with the coefficients in Eqs.~\eqref{eq:fit_coeff} (cyan dashed) and~\eqref{eq:apfit_coeff} (orange dot-dashed), as well as the Tolman VII model in Eq.~\eqref{eq:alpha-C-Tol}. Observe that the second fit was constructed directly from the $ \langle  \alpha \rangle_p$--$C$ data presented here, while the first fit was built from the $y_*$--C data to effectively account for the higher compactness corrections that were ignored in Eq.~\eqref{eq:y_PN}, which is only valid in the Newtonian case.
}
\end{figure}

Figure~\ref{fig:alpha-C} presents the above relation between $ \langle  \alpha \rangle_p$ and $C$ for the Tolman VII model and compares it with the fit in Eq.~\eqref{eq:a_fit_summary}. We also show $\langle  \alpha \rangle_p$ for selected EoS samples. Observe that the Tolman VII relation matches well with the numerical data in the intermediate compactness regime while the fit is slightly off. We stress that the fit was \emph{not} constructed from this $\langle  \alpha \rangle_p$--C data, but rather, it was constructed from the $y_*$--C data. This allows us to effectively take into account higher-order corrections in compactness that were ignored in Eq.~\eqref{eq:y_PN}. 

To emphasize this point further, let us now show how bad the fit becomes if one uses $\bar \alpha$ in the Newtonian, polytrope-inspired $y_*$ (Eq.~\eqref{eq:y_poly_insp}) that more accurately reproduces the data points in Fig.~\ref{fig:alpha-C}. We first construct a quadratic fit for $\langle \alpha \rangle_p$ against $C$ as in Eq.~\eqref{eq:a_fit_summary}, using the data in Fig.~\ref{fig:alpha-C}. Doing so, we find the coefficients
\begin{equation}
\label{eq:apfit_coeff}
    (a_0,a_1,a_2) = (1.35,-9.91,44.2)\,.
\end{equation}
and this fit is also shown in Fig.~\ref{fig:alpha-C}. We next use this fit for $\bar \alpha$ in $y_*$ of Eq.~\eqref{eq:y_poly_insp} and then insert this into Eq.~\eqref{eq:Lambda_C} to find a new, Newtonian, polytrope-inspired Love-C relation.

\begin{figure}
\includegraphics[width=8.5cm]{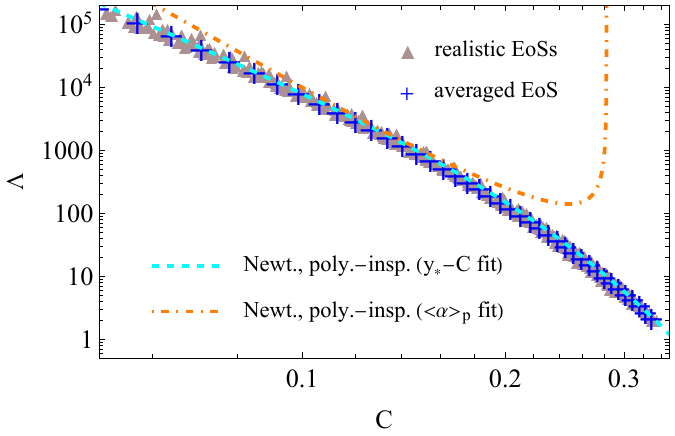}
\caption{\label{fig:Love-C_alphafit}
Similar to the top panel of Fig.~\ref{fig:summary}, but here we compare only the original Newtonian, polytrope-inspired Love-C relation (cyan dashed) against a new one (orange dot-dashed). The former is obtained from the $\bar \alpha$--$C$ quadratic fit (Eq.~\eqref{eq:a_fit_summary} with the coefficients in Eq.~\eqref{eq:fit_coeff}) found by fitting the $y_*$--$C$ data, while the latter is derived from directly fitting the $\langle \alpha \rangle_p$-C data shown in Fig.~\ref{fig:alpha-C} with a quadratic fit (Eq.~\eqref{eq:a_fit_summary} and the coefficients in Eq.~\eqref{eq:apfit_coeff}). Observe that the new relation is not a good model with an unphysical divergence at large compactnesses.
}
\end{figure}

Figure~\ref{fig:Love-C_alphafit}  compares the original Newtonian, polytrope-inspired Love-C relation from Fig.~\ref{fig:summary} to the new one found through the $\langle \alpha \rangle_p$--$C$ fit.  
It is obvious that this new fit is a bad model, especially in the high compactness regime, where the new fit has an unphysical divergence. For assurance, we also tested cubic and quartic fits for $\langle \alpha \rangle_p$ versus $C$, which ultimately resulted in the same qualitative behavior. The reason is, again, that the expression for $y_*$ in Eq.~\eqref{eq:y_poly_insp} is inspired by a \textit{Newtonian} polytrope and the model becomes inconsistent in the high compactness regime if one tries to account for the relativistic corrections only in $\bar \alpha$.

\newpage

\bibliography{master}
\end{document}